# K-núcleo: Una herramienta para detectar la estructura conceptual de los campos de investigación. El caso práctico de la Altmetría

Carmen Gálvez[1]
*Universidad de Granada*

## RESUMEN

En el Análisis de Redes Sociales (ARS), la descomposición k-núcleo se utiliza para la detección de las capas jerárquicas en las redes complejas. La aplicación de la medida k-núcleo a una red de palabras-clave permite representar la estructura conceptual de un campo de investigación. El objetivo de este trabajo fue proponer la aplicación de la métrica k-núcleo para mostrar la jerarquía temática y la evolución de la estructura conceptual del campo de investigación de la Altmetría. La metodología se desarrolló en las siguientes fases: recopilación de datos, selección de palabras-clave, elaboración de una matriz de co-ocurrencia de palabras-clave, generación de una red de palabras-clave, descomposición k-núcleo y visualización de la estructura jerárquica. El resultado fue la detección de cinco capas diferenciadas. Una capa central con conceptos básicos y densamente interconectados, que formaron la base de conocimiento del campo. Una capa intermedia con conceptos mediadores, que mostraron la evolución del conocimiento en el campo. Una capa lateral con conceptos que indicaron la especialización del campo de investigación. Una capa borde con conceptos periféricos y aislados, que representaron los frentes conceptuales en vías de desarrollo. La conclusión fue que la división jerárquica de la red de palabras-clave logró una comprensión más profunda de la estructura conceptual del campo de investigación analizado.

## ABSTRACT

In Social Network Analysis (SNA), k-core decomposition is used to detect hierarchical shells in networks. The application of the K-core decomposition to a network of keywords allows us to represent the conceptual structure of a research field. The objective of this work was to propose the application of k-core decomposition to show the evolution of the conceptual structure of the Altmetrics research field. The methodology was developed in several phases: data collection, keyword selection, elaboration of a keyword co-occurrence matrix, generation of a keyword network, k-core decomposition and visualization of the hierarchical structure. The result was the detection of five differentiated shells. A core shell with basic, densely interconnected concepts that formed the knowledge base of the field. An intermediate shell with mediating concepts that showed the evolution of knowledge in the field. A lateral shell with concepts that indicated the specialization of the research field. A border shell with peripheral and isolated concepts, which represented the conceptual fronts in development. In conclusion, the hierarchical decomposition of the keyword network achieved a deeper understanding of the conceptual structure of the research field.

[1]*Contacto con los autores: Carmen Gálvez (cgalvez@ugr.es).*

# INTRODUCCIÓN

El Análisis de Redes Sociales (ARS) se considera una metodología para el estudio de las relaciones entre diferentes actores, a través de redes compuestas por nodos y enlaces. El ARS ha demostrado ser una herramienta muy valiosa para describir y cuantificar sistemas complejos en muchas ramas de la ciencia (Clauset et al., 2008). Las redes a menudo exhiben una organización jerárquica, en la que los vértices se dividen en grupos que se subdividen a su vez en otros grupos, y así sucesivamente en múltiples fracciones. La jerarquía se presenta como un factor central para explicar muchos fenómenos que se producen dentro de las redes complejas. Esta característica se puede utilizar para analizar la estructura conceptual y temática de los campos y áreas de investigación.

A su vez, para identificar la estructura conceptual de los campos de investigación se han desarrollado métodos cuantitativos, tales como el análisis de palabras asociadas, o análisis de co-palabras. El método de las palabras asociadas se basa en el cómputo de las apariciones conjuntas de palabras, que se obtienen a partir de la frecuencia de aparición en los documentos de un fichero (Callon et al., 1986). El análisis de co-palabras constituye un instrumento útil para descubrir la estructura conceptual y temática de un dominio de conocimiento o un campo específico de investigación (Van Raan, 2005). Los resultados del análisis de co-palabras se pueden visualizar en diagramas estratégicos y en redes de palabras aplicando técnicas propias de ARS (Carrington et al., 2005). Una de las ventajas de la representación del análisis de co-palabras con técnicas procedentes del ARS es la posibilidad de sondear las redes complejas de manera detallada, mediante la identificación de estructuras dentro de la red (como grupos de nodos estrechamente conectados en comunidades de nodos).

Las redes complejas se pueden descomponer en comunidades si los nodos de dicha red pueden ser agrupados según la aplicación de diferentes medidas. Un par de nodos tiene mayor probabilidad de estar conectado si ambos son miembros de la misma comunidad, y menor probabilidad de estar conectados a otros. Uno de los problemas fundamentales en la aplicación de técnicas de detección de comunidades es definir de manera cuantitativa una comunidad en un grafo. De manera intuitiva, podríamos definir una comunidad como aquel conjunto de nodos interconectados por una mayor cantidad de aristas que los demás nodos (Fortunato, 2010).

El objetivo de este trabajo fue la aplicación de medidas de detección de comunidades a redes complejas de palabras-clave para identificar la estructura conceptual y temáticas de los campos de investigación y mostrar su evolución. Esta aproximación se aplicó al caso práctico del campo de la Altmetría, pero podría extraporlarse a otras áreas científicas.

# MARCO CONCEPTUAL

## Detección de comunidades en redes complejas de palabras

Las redes complejas de palabras, o redes lingüísticas, se basan en los contenidos de títulos, resúmenes, textos completos o diversos tipos de información controlada por términos (palabras-clave), para analizar la proximidad semántica, conceptual y temática. Los métodos de análisis de co-ocurrencias de palabras van desde los puramente léxicos hasta los semánticos. Los procedimientos léxicos suelen ser más rápidos, automáticos y menos dependiente del campo analizado, los procedimientos semánticos son más precisos, pero también más difícil de realizar. Las asociaciones de términos han sido ampliamente estudiadas por sociólogos de la ciencia para caracterizar las áreas de conocimiento (Zitt & Bassecoulard, 2008). El análisis de co-ocurrencia de palabras, en combinación con las técnicas de ARS, consigue crear mapas reticulares que permiten la detección de estructuras subyacentes, la visualización de la información y la transferencia de conocimiento.

El análisis de co-ocurrencisa de palabras se puede realizar a partir de un conjunto de documentos representativos de la producción de un área científica. Esta técnica parte del principio según el cual una especialidad de investigación puede ser identificada por su propio vocabulario. La relación de co-ocurrencia se da entre dos elementos que aparecen conjuntamente en un documento, es decir, si el elemento *i* y *j* aparecen en el mismo documento, se dice que existe una relación de co-ocurrencia entre ellos. Las relaciones entre las unidades de análisis pueden representarse como un grafo o una red, donde las unidades suelen ser los nodos y las relaciones entre ellas se representan mediante enlaces entre los nodos. En la primera fase, con las palabras seleccionadas, de una muestra obtenida de forma automática en una base de datos, se construyen matrices de co-ocurrencias de palabras. Una vez obtenida las matrices, se aplican diferentes tipos de análisis, de tal forma que la medida del enlace entre dos palabras será proporcional a la co-ocurrencia de esas dos palabras en el conjunto de documentos que se ha tomado como muestra.

A partir del cómputo de las apariciones conjuntas de palabras, representadas como relaciones, se construyen redes bibliométricas de palabras y mapas científicos que muestran la estructura conceptual y temática del campo de investigación analizado. La creación de los mapas científicos se realiza empleando dos procedimientos a las unidades de análisis: 1) métodos para identificar las relaciones de asociación, aplicando algoritmos de agrupamiento (*clustering*), consistentes en descomponer la red de palabras en grupos similares e interconectados; y 2) métodos para identificar las relaciones jerárquicas, aplicando técnicas procedentes del ARS y técnicas de minería de grafos (*graph mining techniques*), consistentes en la caracterización de las relaciones estructurales entre las subredes, o conjuntos de nodos altamente conectados.

## Relaciones de asociación y técnicas de *clustering*

La identificación de las relaciones de asociación, a través de un análisis de agrupamiento (o *clustering*), consistente en descomponer la red de co-palabras en grupos similares e interconectados. Las asociaciones entre las palabras obtenidas serían asimilables a las líneas temáticas o frentes de investigación de los campos científicos analizados. Las técnicas de agrupamiento dividen un conjunto de elementos en diversos subconjuntos, los cuales deben cumplir la condición de tener una gran cohesión y similitud interna. Los diversos algoritmos de *clustering* aplicados a las redes bibliométricas intentan descubrir las subredes que forman la red bibliométrica global, es decir, aquellos conjuntos de nodos que están fuertemente enlazados entre sí, pero pobremente enlazados con el resto de la red. La aplicación de las técnicas de *clustering* a las redes bibliométricas de palabras se han utilizado con éxito en diferentes estudios (Callon et al., 1986; Callon et al., 1991; Cobo et al., 2011; Dehdarirad et al., 2014; Ding et al., 2001; Jun & Weilin, 2024; Leydesdorff & Welbers, 2011; Liu et al., 2012; López-Fraile et al., 2023; Luo et al., 2022; Raeeszadeh et al., 2018; Sedighi, 2016; Segado-Boj et al., 2023; Sonni et al., 2024; Xie, 2015).

## Relaciones jerárquicas y técnicas de ARS

El ARS permite construir grafos (o cientogramas) en los que cada palabra se describe por un nodo (vértice) y las líneas entre los nodos representaron las relaciones entre dos palabras. Una de las características de las redes complejas es su estructura comunitaria, es decir, la existencia de grupos de nodos mucho más conectados entre sí que con el resto de la red. Las comunidades reflejan las relaciones topológicas entre elementos del sistema subyacente de la red. En el ARS existe un orden y organización de los vértices, a esta característica se le denominó estructura de comunidad (Wasserman & Faust, 1994). Desde una perspectiva local, un grupo, o comunidad, es un conjunto con la mayor conexión entre los nodos, hasta el punto que se puede analizar de manera independiente del resto de la red (Fortunato, 2010). La detección de comunidades locales en las redes se fundamenta en la búsqueda de conjuntos de nodos altamente conectado (Papadopoulus et al., 2012).

Una comunidad es un grupo de vértices que comparten propiedades comunes y tienen funciones similares dentro del grafo. Las comunidades locales se focalizan en el subgrafo a estudiar ignorando el resto del grafo. La mayoría de las redes reales suelen contener partes en las que los nodos están más conectados entre sí que con el resto de la red. Los conjuntos de tales nodos generalmente se llaman comunidades, grupos cohesivos o módulos (Hanneman, 2000).

Otra característica común de muchas redes con topografía compleja es que las conectividades de los vértices siguen una distribución denominada Redes Invariantes por Escala (RIE) o redes sin escala porque no presentan una conectividad característica (como son las redes aleatorias) (Barabási & Albert, 1999): se dice que una red es sin escala, o de libre escala (*scale-free*) cuando la función de distribución del número de vínculos de los nodos de dicha red sigue una ley de potencias (*power-law*), esto significa que se pueden encontrar pocos nodos que tienen muchos vínculos, y muchos nodos que poseen pocos vínculos. Esta característica es una consecuencia de dos mecanismos: 1) las redes se expanden continuamente mediante la adición de nuevos nodos; y 2) los nuevos nodos se unen preferentemente a nodos que ya están bien conectados. Un modelo basado en esto dos mecanismos reproduce las distribuciones sin escala observadas, lo que indica que el desarrollo de grandes redes se rige por fenómenos que van más allá de los detalles de los sistemas individuales (Barabási & Albert, 1999).

También en el ARS, las estructuras que subyacen a muchas redes presentan una serie de propiedades comunes, como la propiedad de mundo pequeño (*small-world*) (Newman et al., 2003). Una red es un mundo pequeño cuando existen pocos pasos de separación por término medio entre dos nodos, independientemente del tamaño de dicha red (Milgram 1967; Watts & Strogatz 1998). Sin embargo, el avance de los métodos matemáticos y el desarrollo de nuevas herramientas analíticas ha permitido un cambio de perspectiva para el análisis de las redes grandes y complejas. Con este cambio de perspectiva, el ARS se enfrentó a nuevos desafíos (Newman, 2003): 1) el análisis ha pasado de redes pequeñas (con pocos nodos y sencillas de explorar en grafos por el ojo humano) a redes complejas (con miles de nodos y muy difíciles de explorar a simple vista); 2) el foco de atención ha ido cambiando desde las propiedades individuales de los nodos al análisis de las propiedades estadísticas y estructurales a gran escala de las redes; y 3) la muestra de datos ha trascendido del análisis tradicional con muestras de tamaño reducido (con recogida de datos a través de encuestas y entrevistas, que limitan el tamaño de la red), al tratamiento masivo de datos, obtenidos de forma automática, como son las muestras recuperadas en bases de datos.

Teniendo en cuenta las características de las redes complejas para explicar los diversos fenómenos de los sistemas complejos, en la detección de los grupos, o subgrafos, hay una serie de métricas dentro de la minería de grafos para encontrar un patrón o una relación en un grupo de gafos. Entre las métricas basadas en el principio de la identificación de los nodos más densamente conectados, se encuentran: *cliques*, *n-cliques* y *k-cores* (Wasserman & Faust, 1994).

En relación a la métrica *k-core*, dado un grafo no-dirigido, un k-núcleo (*k-core*) (Alvarez-Hamelin et al., 2005) es el subgrafo máximo en el que todos sus nodos están conectados con al menos otros *k* puntos (Dorogovtsev et al., 2006). La descomposición k-núcleo fue utilizada por primera vez para medir la densidad local y la cohesión en redes sociales (Seidman, 1983). El algoritmo k-núcleo pronto se convirtió en uno de los métodos más populares para detectar la estructura de las redes debido a su simplicidad y aplicabilidad. La métrica k-núcleo consiste básicamente en descomponer la red de nodos en núcleos cada vez más cohesionados, a través de un proceso sucesivos anidamientos, que identifican subconjuntos cada vez más cohesionados. Según los valores de conexión de los nodos, la estructura de una red se descompone en núcleos y capas (*shells*) jerárquicas, desde las capas externas a las más centrales, que se representan gráficamente en cajas que se contienen entre ellas, similar al anidamiento de una muñeca rusa (Figura 1).

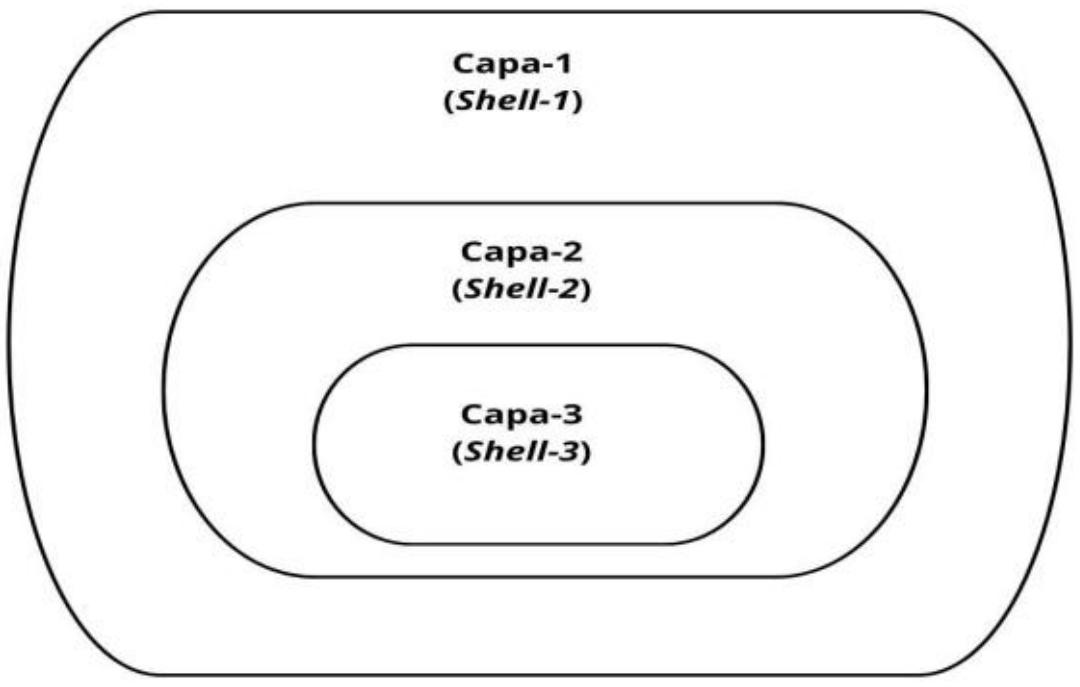

**Figura 1**. Estructura de red en capas (*shells*) jerárquicas. Fuente: Elaboración propia.

Siguiendo con lo anterior, la métrica para realizar la descomposición k-núcleo se fundamenta en un algoritmo de poda (*pruning*) que elimina recursivamente los nodos que tengan grados menores que *k*. Para ser incluida en el k-núcleo, un nodo debe estar vinculado al menos a otros *k* ítems del grupo. El núcleo cero (*k=0*) equivaldría al grafo completo. A medida que *k* aumenta, los tamaños de los núcleos disminuyen y los núcleos se vuelven más interconectados. Los k-núcleos con mayores valores representan las regiones más cohesivas del gráfico y las zonas más centrales de la red. El resultado ofrece una visión organizada en una jerarquía de k-capas (*k-shells*), con los nodos más centrales en la capa de mayor *k* (Figura 2).

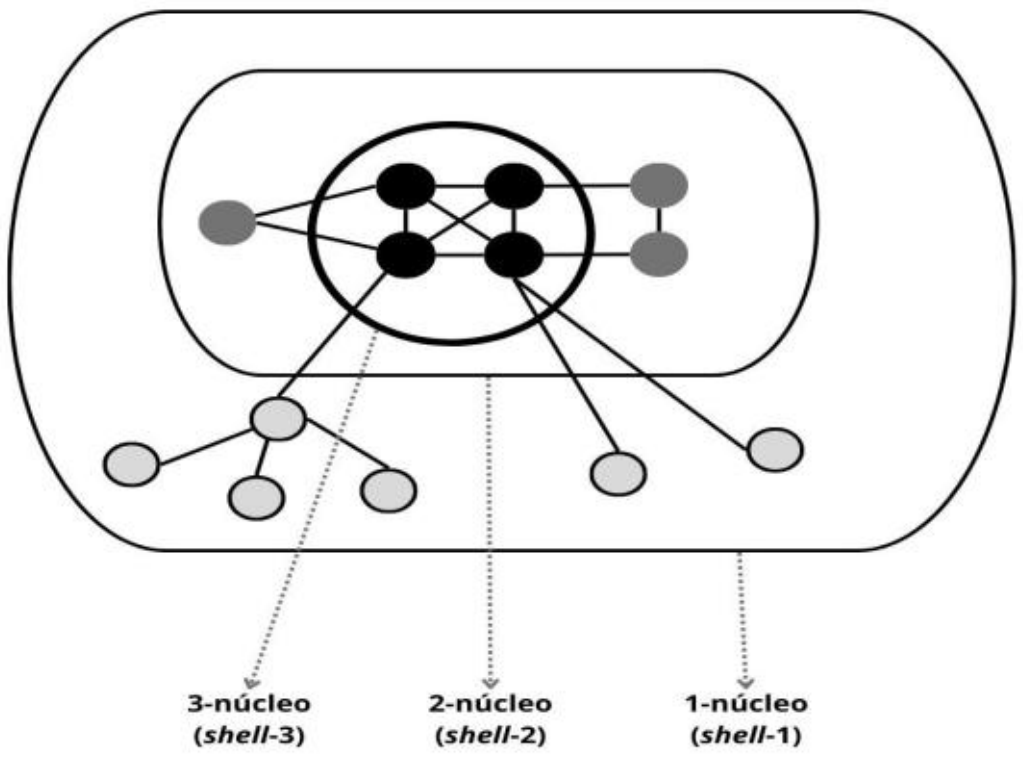

**Figura 2**. K-núcleo de una red. Fuente: Adaptacion k-núcleo para una red pequeña (Alvarez-Hamelin et al., 2005).

La detección de comunidades constituye una herramienta valiosa para el análisis de redes complejas, al permitir el estudio de estructuras que a menudo están asociadas con características organizativas y funcionales de las redes subyacentes. Descubrir subgrafos densos en un grafo es una tarea fundamental de minería

de grafos (Li et al., 2018). La descomposición k-núcleo se ha aplicado a redes del mundo real, como Internet (Alvarez-Hamelin et al., 2008) o redes celulares (Wuchty & Almaas, 2005). Así mismo, la detección de comunidades ha demostrado ser útil en una serie de dominios, como la Biología, las Ciencias Sociales o la Bibliometría (Padopoulus et al., 2012). Diversos estudios han proporcionado evidencia de que las redes del mundo real a menudo están organizadas jerárquicamente (Alvarez-Hamelin et al. 2005; Clauset et al. 2008; Du et al., 2007; Fortunato, 2010; Ravasz et al. 2002; Sales-Pardo et al., 2007; Yang & Dia, 2008). Este algoritmo se adoptó para encontrar las partes más densas de una red (Kong et al., 2019; Liu & Zhong, 2024; Malvestio et al., 2020).

En la actualidad, la combinación del método k-núcleo, junto con el análisis de co-palabras, se ha utilizado en la construcción de mapas científicos en diversas áreas de conocimiento tales como la evolución de la investigación en el campo de los sistemas de gestión de información (Choi et al., 2011), la investigación de bibliotecas digitales chinas (Xiao et al., 2016), estudios sobre nuevos indicadores bibliométricos (Wang & Chai, 2018), los estudios sobre la gestión de la crisis ambiental (Dai et al., 2020) o en el análisis de la depresión en el campo de la psicología (Yu et al., 2022). A pesar de estos trabajos previos, se ha prestado poca atención a la detección de comunidades en las redes complejas de palabras para explorar la estructura jerárquica de los campos de investigación.

## El caso práctico del campo de la Altmetría

La Altmetría ha provocado una transformación en la evaluación de la información científica, debido a la difusión de la ciencia en los nuevos medios sociales y a la irrupción del movimiento acceso abierto (*open access*). La publicación del *Manifiesto Altmetrics* (Priem et al., 2011) supuso el inicio de la superación de las limitaciones de los indicadores bibliométricos tradicionales. Las altmétricas engloban todas aquellas métricas alternativas que miden las nuevas formas de ejercitar, discutir o comunicar la ciencia, especialmente en las plataformas de contenidos generados por los usuarios. Los indicadores altmétricos (o métricas de los nuevos medios derivados de las webs sociales) (Thelwall, 2017) se basan en la información proveniente de diversas fuentes (tales como, blogs, medios de comunicación y redes sociales), en los que se mide la participación de los usuarios en la web social, con fines académicos. Los términos web 2.0 y web social se refieren a aquellos sitios web que facilitan el uso de la información, la interoperabilidad, el diseño centrado en el usuario y la colaboración. Este cambio ha provocado que la mayor parte de los investigadores hayan trasladado sus actividades de investigación a la web y a los medios sociales. Además, esta situación se ha hecho más evidente debido a que actualmente se dispone de herramientas que tienen más potencialidades para desarrollar un rango mayor de influencia académica que los entornos tradicionales de publicación.

Las métricas alternativas o altmétricas basadas en las redes sociales (para distinguirlas de la bibliometría) se consideran una opción interesante para evaluar el impacto social de la investigación (Bornmann, 2014). Las altmétricas engloba métricas basadas en la web para el impacto del material académico, con énfasis en los medios de comunicación sociales como fuentes de datos (Shema et al., 2014). La importancia de esta forma alternativa de métricas quedó reflejada por uno de los mayores proveedores de bases de datos multidisciplinarias, como es Elsevier, que compró Mendeley (una aplicación web que permite gestionar referencias bibliográficas y documentos de investigación, combinando un administrador de citas con una red social académica) (Roemer & Borchardt, 2013). Los investigadores empezaron a incluir indicadores altmétricos, junto con la medición de impacto de las citas, en sus publicaciones (Piwowar y Priem, 2013). Comenzó a hablarse de un nuevo paradigma de evaluación de la investigación (Lin & Fenner, 2013).

La Altmetría se constituyó como un área dinámica, con muchos desafíos y de rápido crecimiento (Galloway et al., 2013). Las altmétricas han recibido mucha atención por parte de los académicos en los últimos años. Como campo de investigación emergente en ciencias sociales, ha recibido mucha atención de los académicos (Cho, 2021; Galloway et al., 2013; Liu et al., 2020; Tahamtan & Bornmann, 2020). Se trata de un área en expansión, que ha dado lugar a nuevas alternativas de evaluación de la actividad científica. En este trabajo, se aplicó el método de la descomposición k-núcleo al campo específico de la Altmetría para identificar y visualizar la estructura jerárquica de los conceptos y temas de investigación que aborda este campo.

## MÉTODO

Una vez recopilados los datos, la investigación se desarrolló en las siguientes etapas básicas (Figura 3): 1) selección de las palabras-clave; 2) creación de una matriz de co-ocurrencia de palabras-clave; 3) generación de una red no-dirigida de palabras-clave; 4) descomposición k-núcleo; y 5) visualización e interpretación de la estructura jerárquica resultante. La combinación de un conjunto de herramientas informáticas ha facilitado este procedimiento. Para la construcción de la matriz de co-ocurrencia de palabras-clave se utilizó el programa *Bibexcel* (Persson, Danell & Wiborg Schneider, 2009), que permitió la preparación de los datos y su posterior volcado a un programa de ARS. Para la obtención de la red de palabras-clave y la partición k-núcleo se utilizaron herramientas específicas de ARS: *Ucinet* y *NetDraw* (Borgatti, 2002). Para la visualización de la red de palabras-clave resultante se empleó el software *VOSviewer* (Van Eck & Waltman, 2010).

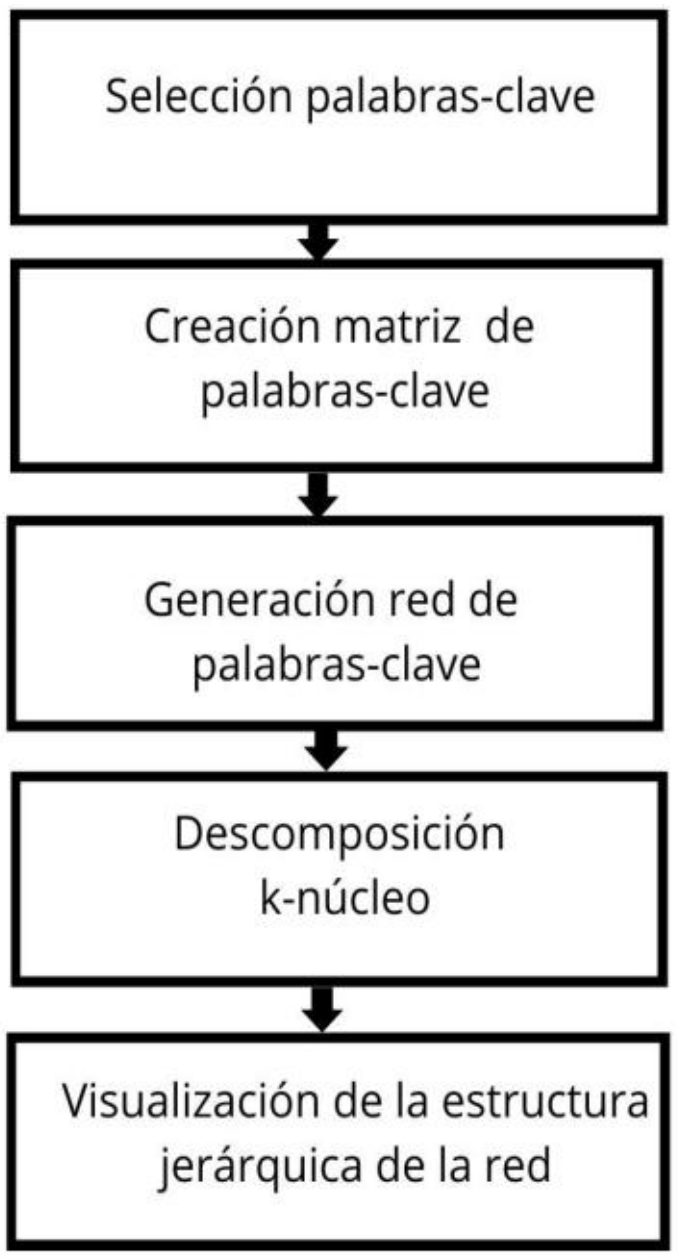

**Figura 3.** Etapas básicas de la metodología empleada. Fuente: Elaboración propia.

## Recopilación de datos

La muestra de datos se obtuvo de la plataforma *Web of Science* (WoS), propiedad de la empresa *Clarivate Analytics*, compuesta a su vez por una colección de bases de datos de referencias bibliográficas y citas de publicaciones periódicas. Dentro de WoS se seleccionaron las siguientes bases de datos: *Science Citation Index Expanded (*SCIE), *Social Sciences Citation Index* (SSCI) y *Arts & Humanities Citation Index* (A&HCI). La elección de WoS estuvo motivada por tres criterios: 1) recoge las referencias de las principales publicaciones científicas de cualquier disciplina del conocimiento a nivel internacional; 2) el tipo de indización que utiliza WoS, los documentos se indizan por palabras-clave (*Keywords Plus*), generadas de forma automática por el sistema, y palabras-clave de autor (*Author's Keywords*), o términos proporcionados por los propios autores, mucho más específicos que los términos generados de forma automática; 3) permite la exportación de los datos en diferentes formatos, lo que posibilita la reestructuración y el posterior procesamiento en diferentes aplicaciones analíticas necesarias para el tratamiento en herramientas de ARS.

La estrategia de búsqueda avanzada empleada consistió en utilizar los términos "altmetric" y "altmetrics" en el campo "*Topic*" de WoS y "*Article*" en el campo "Tipo de documento". El periodo de búsqueda fue: 2017-2024. La consulta se realizó el 1 de mayo de 2024. Los documentos recuperados fueron descargados directamente en formato de texto plano para ser procesados.

## Selección de palabras-clave

Se seleccionaron la palabras-clave de autor por ser más específicas que las palabras-clave generadas de forma automática por WoS. A continuación, se probaron distintos umbrales de frecuencias de palabras-clave para seleccionar sólo un límite de términos relevantes. Se observó que los resultados más claros se obtuvieron cuando se consideraron sólo aquellas palabras-clave cuya frecuencia fuera ≥4 ocurrencias en la muestra analizada. Si se hubieran seleccionado las palabras-clave con una frecuencia menor, la red resultante posterior hubiera sido difícil de interpretar por la gran cantidad de nodos y enlaces obtenidos. También se eliminaron manualmente las palabras-clave '*altmetric*' y '*altmetrics*' (porque presumiblemente se supuso que estarían relacionadas con todos los demás términos de la red). Es necesario aclarar que las palabras-clave '*altmetric*' y '*altmetrics*' se eliminaron cuando aparecieron solas, pero no se excluyeron en los casos en los que formaron parte de términos compuestos, como '*altmetric attention score*'.

## Creación de una matriz de palabras-clave

Con las palabras-clave de autor seleccionadas se generó, de forma automática, una matriz de P x P elementos (palabras-clave por palabras-clave), utilizando la aplicación *Bibexcel* (Persson et al., 2009): donde P es la palabras-clave a representar, a partir de las veces que un par de palabras-clave parece conjuntamente en los documentos. Cuanto más co-existieron dos palabras-clave, más estrecha se consideró su relación. Las características de la matriz de palabras-clave fueron las siguientes: 1) adyacente, o cuadrada, mismo número de términos en filas y columnas; 2) modo-1, los mismos términos en filas y columnas; 3) simétrica, el mismo valor en las filas y las columnas; 4) binarias, se asignó el valor "1" cuando se produjo una co-ocurrencia de dos palabras-clave en los documentos, y valor "0" si no se produjo.

### Generación de una red de palabras-clave

Partiendo de la matriz de co-ocurrencias, se construyó un grafo (o cientograma), con *NetDraw* (Borgatti, 2002), en el que cada palabra-clave se describió por un nodo (vértice) y la relación entre los nodos se representó por líneas, el grosor de las líneas indicó la intensidad de las relaciones entre dos palabras-clave.

### Descomposición k-núcleo

Para detectar los grupos, o subgrafos de nodos, dentro de las estructuras de la red compleja de palabras-clave, se utilizó la partición k-núcleo (*k-core*) con la herramienta *NetDraw* (Borgatti, 2002). El algoritmo para realizar la descomposición k-núcleo consistió en un proceso de poda que eliminó recursivamente los nodos que tuvieran grados menores que k. Para ser incluido en el k-núcleo, un nodo debió estar vinculado al menos a otros *k* nodos del grupo. A medida que *k* aumentó, los núcleos se volvieron más interconectados y cohesionados.

### Visualización e interpretación de la estructura jerárquica resultante

La red de palabras-clave resultante se representó en una estructura jerárquica. Los mayores valores representaron las regiones más cohesivas del gráfico y las zonas más centrales de la red. Las zonas con menores valores se situaron en las zonas periféricas de la red. La estructura jerárquica de los núcleos mostró la estructura temática y la evolución del campo de investigación de la Altmetría. Para permitir una mejor visualización e interpretación, la red de palabras-clave obtenida se visualizó en un mapa bidimensional (2D) etiquetado utilizando la aplicación *VOSviewer* (Van Eck & Waltman, 2010), en el que cada palabra-clave estuvo representada por una etiqueta y un círculo cuyo tamaño fue proporcional al grado de intensidad de las relaciones internas de los nodos que conformaron la red.

## RESULTADOS

Con los criterios de búsqueda seleccionados, se obtuvo un total de 983 registros, con 2.017 palabras-clave únicas asignadas por los autores. Después de seleccionar las palabras-clave con una frecuencia ≥4, la muestra se redujo a 228 palabras-clave (Tabla 1). Con los términos seleccionados se construyó, de forma automática, una matriz de 228 x 228 palabras-clave. A partir de esta matriz se obtuvo una red de palabras-clave, en la que cada término se describió por un nodo (o vértice), y el grosor de las líneas, entre los nodos, representaron la intensidad de las relaciones entre las palabras-clave. A continuación, se aplicó la descomposición k-núcleo a la red de palabras-clave.

El resultado fue una red de 228 nodos y 13 núcleos, que revelaron la estructura semántica y temática del campo de investigación de la Altmetría (Figura 4). En la estructura jerárquica, los núcleos con mayores valores representaron las capas (*shells*) más cohesivas del gráfico y las zonas de investigación más centrales de la red. Cuanto más intensas fueron las relaciones entre palabras-clave, más constituyeron un conjunto coherente e integrado de frentes temáticos. Cuando estas relaciones fueron débiles, los agregados se redujeron a conexiones entre palabras-clave que no se cerraron los unos con los otros, dando lugar a focos menos relevantes, o en vías de desarrollo.

**Tabla 1**

*Información global de los datos obtenidos de Web of Science (WoS).*

| Descripción | Resultados |
|---|---:|
| Período | 2012-2024 |
| Artículos | 983 |
| Tasa anual de crecimiento % | 20,46 |
| Palabras-clave (*Keyword Plus*) | 993 |
| Palabras-clave de autor (*Author's Keywords*) | 2.017 |

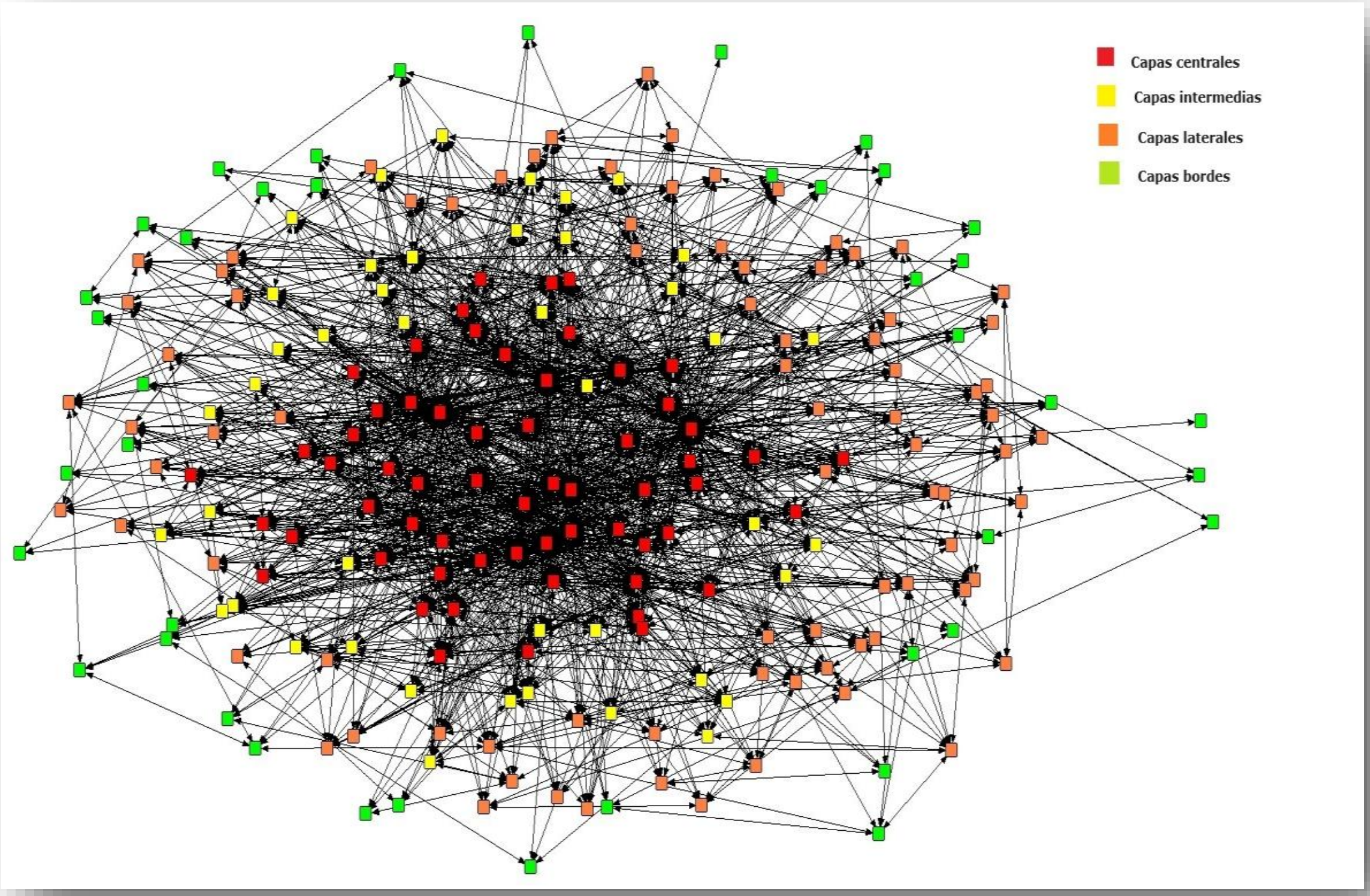

**Figura 4.** Aplicación de la métrica k-núcleo a la red de palabras-claves y descomposición k-núcleo en capas jerárquicas. Fuente: Elaboración propia.

Las capas resultantes, según la métrica k-núcleo, se dividieron en: central, intermedia, lateral y borde. Debido a que muchas capas adyacentes mostraron una estructura similar se decidió fusionarlas para simplificar su análisis. El resultado de la aplicación de la métrica k-núcleo (del centro a la periferia) fue el siguiente (Tabla 2):

- Capa central (*k=13* y *k=12*). Incluyó 58 conceptos básicos fuertemente, interconectados que conformaron la base de conocimiento del campo.
- Capa intermedia *k=11*, *k=10* y *k=9*). Incluyó 42 conceptos mediadores, directamente conectados con la capa básica, que mostraron la evolución del conocimiento en el campo.
- Capa lateral (*k=8*, *k=7* y *k=6*). Incluyó 87 conceptos que indicaron la especialización del campo de investigación.
- Capa borde o marginal (*k=5*, *k=4*, *k=3, k=2* y *k=1*). Incluyó 41 conceptos periféricos y aislados, que representaron los frentes en vías de desarrollo, o en vías de desaparición

La red de palabras-clave dividida en las diferentes capas se representó en un mapa bidimensional (2D) (Figura 5), utilizando el programa de visualización de redes VOSviewer (Van Eck & Waltman, 2010). En la red, los grupos de palabras-clave se representaron por un círculo y se identificaron por una etiqueta, con el término de mayor conexión dentro del grupo de palabras-clave. Cuanto mayor fue el valor de la métrica k-núcleo mayor fue el diámetro del los círculos y las etiquetas. El color aleatorio de los círculos estuvo determinado por la capa a la cual se adscribió cada conglomerado de palabra-clave. Las líneas entre los elementos representaron los enlaces. En el centro del mapa se situaron los grupos de palabras-clave con mayor conexión, y, por tanto, con mayor relevancia científica. En los bordes del mapa se situaron los grupos palabras-clave con menos conexión, y por tanto, menos relevantes.

**Tabla 2**

*Capas jerárquicas (central, intermedia, lateral y borde) obtenidas después de la aplicación de la métrica k-núcleo a la red de palabras-clave. Fuente: Elaboración propia.*

| Núcleo | Capa | Palabras-clave |
|---|---|---|
| *k=13* | Central | *bibliometrics; social media; mendeley; twitter; facebook; scholarly communication; open access; scientometrics; research evaluation; altmetric.com; citation; google scholar; analysis; citations; covid-19; dimensions; h-index; impact; impact factor; mendeley readership; metrics; news; online attention; research impact; researchgate; scientific communication; scientific impact; scientific publication; scopus; social media metrics; social networks; social web; societal impact; visibility; web of science; wikipedia.* |
| *k=12* | Central | *social network analysis; plumx; open science; communication; dissemination; evaluation; humanities; indicators; information; journal impact factor; library and information science; scholarly impact; science communication; science mapping; scientific journals; scientific production; social impact.* |
| *k=11* | Intermedia | *cybermetrics; informetrics; webmetrics; web indicators; citation impact; gender.* |
| *k=10* | Intermedia | *blogs; book impact; scholarly books; science of science; sentiment analysis; coverage; downloads; information science; machine learning; mentions; publications; spain*. |
| *k=9* | Intermedia | *artificial intelligence; science 2.0; social sciences; universities citation prediction; collaboration; coronavirus; diversity; faculty; peer review; public engagement; public policy; publishing; readership; research dissemination; researchers; scholarly publishing; usage statistics.* |
| *k=8* | Lateral | *institutional repositories; preprints; academia.edu; librarians; librarianship; article-level metrics; co-citation analysis; comparative analysis; content analysis, crossref; databases; digital scholarship; early career researchers; higher education; journals; monographs; orthopaedics; peer-review; science; scientific evaluation; usage metrics; visualization.* |
| *k=7* | Lateral | *link analysis; social attention; information dissemination; conference papers; correlation analysis; factor analysis; highly cited papers; information retrieval; intellectual structure; journal metrics; neurosurgery; quality; readership analysis; research quality; research support; vosviewer.* |
| *k=6* | Lateral | *academic networks; academic evaluation; wos usage; author-level metrics; citation rate; highly cited articles; interdisciplinarity; metric; networks; policy documents; psychology; ranking; research assessment; research metrics; scientific collaboration; social sciences and humanities; statistics; traditional metrics.* |
| *k=5* | Borde | *scholarly communications; university ranking; citation advantage; economics; hashtags; health; journal impact factors; meta-analysis; topic modelling; wechat.* |
| *k=4* | Borde | *scientific misconduct; knowledge diffusion; article usage; career development; data quality; equity; g-index; impact evaluation; india; latin america; prediction; retractions.* |
| *k=3* | Borde | *altmetric attention score; library statistics; mantel-haenszel quotient (mhq); academic libraries; assessment; cardiovascular articles.* |
| *k=2* | Borde | *nature index; rankings; survey.* |
| *k=1* | Borde | *digital media; academic social networking site.* |

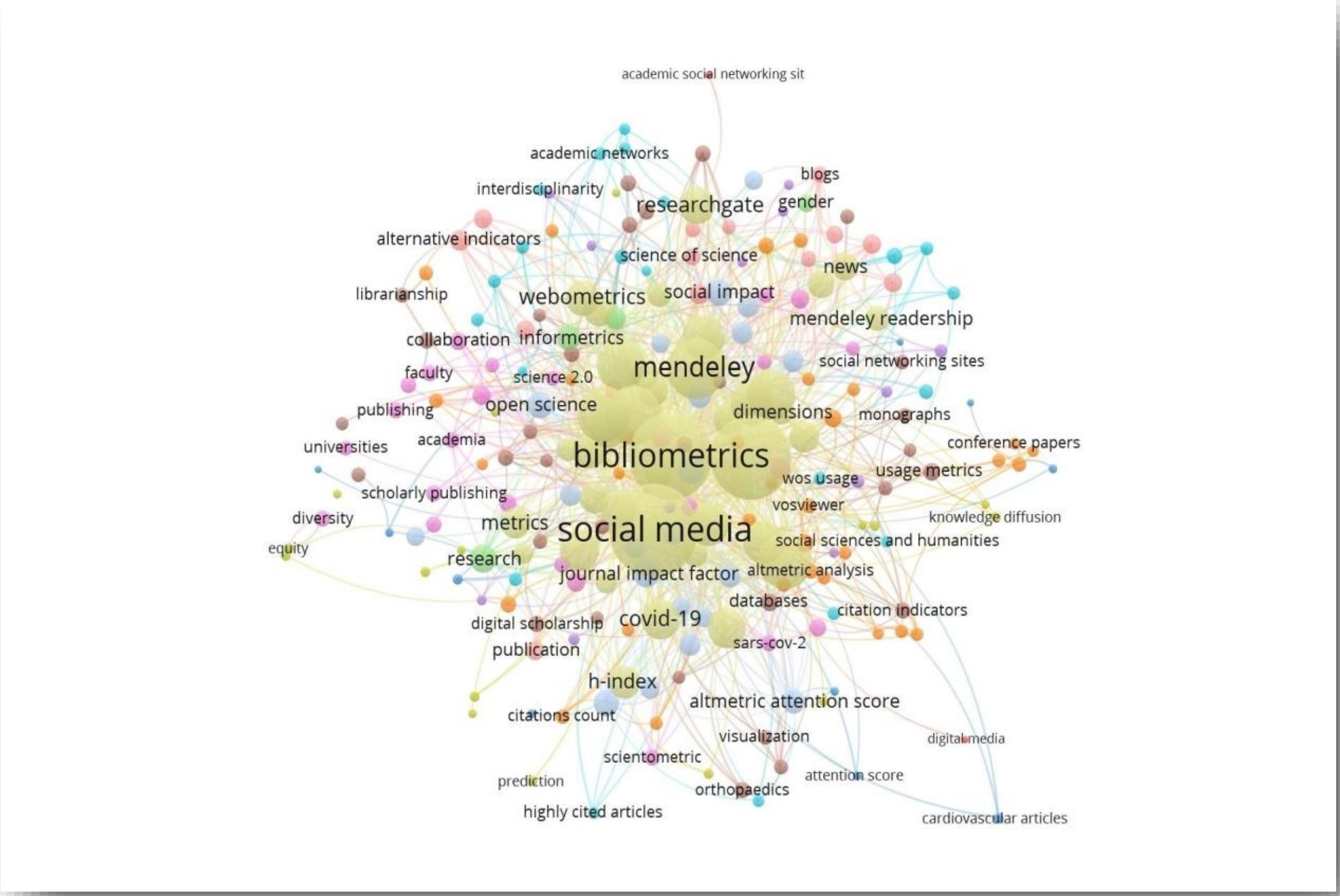

**Figura 5.** Visualización de la red de palabras-clave. Fuente: Elaboración propia.

## DISCUSIÓN

La mayoría de las redes suelen contener partes en las que los nodos están más altamente conectadas entre sí que con el resto de la red. Los conjuntos de tales nodos generalmente se denominan clústeres, comunidades, grupos cohesivos, subgrafos, conglomerados o agregados. La detección de comunidades hace referencia al conjunto de métodos utilizados para identificar subconjuntos de nodos que están muy conectados entre ellos y, al mismo tiempo, poco conectados con el resto de nodos de la red. La identificación de comunidades en redes se realiza para encontrar la estructura jerárquica de los sistemas complejos. Uno de los métodos de identificación de comunidades lo constituye la métrica de minería de grafos k-núcleo. En este trabajo se ha aplicado esta métrica, a través de una combinación de herramientas informáticas de análisis, para detectar la estrutura jerárquica de una red de palabras-clave vinculada al campo de investigación de la Altmetría. Según los valores de esta medida, la estructura de la red de palabras-clave se dividió en núcleos y capas jerárquicas, desde las capas centrales a las más externas.

La morfología de la red de palabras-clave resultante mostró la jerarquía de este campo de investigación dividida en diversos focos temáticos que van desde los más relevantes a los marginales.

En la estructura jerárquica, se identificó un núcleo denso y central vinculado con la correlación entre los indicadores bibliométricos tradicionales y los indicadores altmétricos. Esta capa mostró los conceptos básicos estrechamente interconectados que forman la base de conocimientos del campo. Destacaron palabras-clave como: '*bibliometrics*', '*social media*', '*social network analysis', 'twitter*', '*facebook*', '*scientometrics*'. '*scholary communication*', '*mendeley*', '*open acces*', '*research evaluation*', '*citation analysis*' o '*plumx*'. En esta capa tuvo especial relevancia los medios sociales y las plataformas digitales donde se comparte contenidos por los usuarios. Las redes sociales (tales como, facebook o twitter) se configuraron desde su inicio como la base de estos medios sociales, modificando la forma en la que los usuarios interactúan entre sí, generando y compartiendo información en la web social. En esta capa se situaron las aplicaciones web que se consideran fuentes principales de datos altmétricos, como el gestor de referencias bibliográficas '*mendeley*', la herramienta '*plumx*' que permite analizar el impacto académico de investigadores, o el motor de búsqueda de contenido y bibliografía científico-académica '*google scholar'.* Este núcleo básico también reflejó la iniciativa del acceso abierto ('*open access*'), o acceso ilimitado y gratuito a la información científica y al uso sin restricciones de los recursos digitales.

Se distinguió un núcleo intermedio y transversal (con palabras-clave que conectaron las capas centrales con las capas borde) vinculado con frentes en desarrollo, que mostró la evolución natural del conocimiento en el campo. Destacaron palabras-clave como: '*cybermetrics', 'informetrics', 'webmetrics', 'web indicators', 'blogs', 'artificial intelligence', 'science 2.0', 'social sciences'.* La Cienciometría, forma parte de la sociología de ciencia, se dirige al estudio de los aspectos cuantitativos de la ciencia como disciplina. La Informetría abarca el estudio de los aspectos cuantitativos de la información, independientemente de la forma en que aparezca registrada. También en esta capa se reflejaron los avances en los estudios de la Webometría y Cibermetría, consistes en la aplicación de las técnicas altmétricas al estudio de la relación entre los diferentes sitios de la web social.

Se detectó un núcleo lateral relacionado principalmente con los repositorios digitales y el movimiento del acceso abierto a la ciencia. Esta capa de detalle se compuso de palabras-clave de baja frecuencia cuya semántica es más especializada que las palabras-clave de la capa básica e intermedia. Se distinguieron palabras-clave como: '*institutional repositories', 'preprints', 'academia.edu', 'librarians', 'librarianship', 'article-level metrics' y 'academic networks'.* Esta capa reflejó un avance significativo en campo de los indicadores altmétricos, al incorporarlos a los repositorios digitales de las instituciones y universidades (para obtener datos de estadísticas de uso, descargas y hacer un seguimiento del intercambio de la producción científica). También reflejó la nueva función de los bibliotecários, en las bibliotecas universitarias y especializadas, para gestionar las citas e impacto a sus investigadores.

Por último, se mostró un núcleo borde formado por grupos independientes, con pocos conceptos similares. Las palabras-clave de esta capa mostraron baja frecuencia y pocas conexiones ente ellas. Los conceptos de esta capa marginal evolucionaron a partir de conceptos generales. Destacaron palabras-clave como: '*scholarly communications', 'university ranking', 'scientific misconduct', 'knowledge diffusion', 'nature index' o 'altmetric attention score'.* Esta capa reflejó los posibles temas emergentes, pero todavia no desarrollados, del campo. Debido a su baja frecuencia no se consideraron suficientemente relevantes. Temas como la difusión y divulgación del conocimiento científico (provocado por la creciente demanda de información científica de la sociedad), el tema del fraude científico en las investigaciones dentro de los medios sociales, el indicador alternativo denominado '*Altmetric Attention Score'* (que mide el impacto de un trabajo en función de la repercusión que ha tenido en medios sociales), o la base de datos '*Nature Index'* (que posiciona a los países e instituciones a partir de las contribuciones de investigación publicadas en un grupo de revistas, principalmente pertenecientes a las ciencias naturales).

## CONCLUSIONES

En este trabajo se ha propuesto la métrica k-núcleo para analizar la estructura temáticas y jerárquica de los campos de investigación. Esta metodología se ha aplicado al campo específico de la Altmetría. La topografía de la red mostró cuatro subcampos de interés: 1) un núcleo básico que conectó los indicadores altmétricos con los medios sociales, los gestores de referencias y el movimiento de acceso abierto a los recursos digitales; 2) un núcleo intermedio, que mostró la evolución del núcleo básico, vinculado a la sociología de la ciencia o los aspectos cuantitativos de la información en la web social; 3) un núcleo lateral especializado y emergente vinculado a los repositorios digitales de las instituciones y universidades; y 4) un núcleo marginal, no desarrollado todavía, relacionado con la difusión del conocimiento, el fraude científico, el surgimiento de indicadores altmétricos innovadores y nuevas bases de datos para medir el impacto social de la investigación.

La identificación de comunidades en redes complejas constituye un problema extraordinariamente difícil y para el cual no se ha obtenido una solución que funcione de forma adecuada en cualquier tipo de grafo que pueda presentarse en situaciones reales. Por ello, es crucial desarrollar y contar con herramientas informáticas, algoritmos y métodos que faciliten la tarea de desvelar estas propiedades estructurales de las redes. A pesar de esta dificultad, la aplicación de técnicas de minería de grafos a una red de palabras-clave ha ayudado a distinguir diferencias particulares en la jerarquía semántica de los conceptos y lograr una comprensión más precisa y profunda de un campo de investigación. Este procedimento se podría extraporlar a otras áreas o ámbitos del conocimiento.

Los mapas científicos construidos con técnicas de ARS se pueden utilizar para analizar la evolución de la estructura conceptual de los campos científicos a través de diversos periodos de tiempo. El análisis temporal permite identificar la evolución estructural de los diversos aspectos de un campo científico a través de diferentes periodos de tiempo. En futuros trabajos se pretende realizar un análisis longitudinal, mostrando los aspectos dinámicos de la evolución del campo.

## REFERENCIAS